\documentclass[useAMS,usenatbib]{mn2e}

\usepackage{amsmath}
\usepackage{apjfonts}
\usepackage{amssymb}
\usepackage{mathrsfs}
\usepackage{natbib}
\usepackage{color}
\usepackage{graphicx}
\usepackage{rotate}
\usepackage[normalem]{ulem}

\definecolor{orange}{rgb}{1,0.5,0}

\newcommand{\Msun}{\mbox{$\rm \,M_\odot$}}

\newcommand{\newredshift}{8}

\newcommand{\newmwminit}{7\times10^{9}}
\newcommand{\newsatminit}{2\times10^{8}}
\newcommand{\newsatmfinal}{2\times10^{5}}
\newcommand{\newmwrho}{30}
\newcommand{\newsatrho}{10}

\newcommand{\newsatconcen}{6}
\newcommand{\newecc}{0.9}

\newcommand{\newdecaytime}{13}

\title[Can a Satellite Galaxy Merger Explain
the Active Past of the Galactic Centre?]{Can a Satellite Galaxy Merger Explain
the Active Past of the Galactic Centre?}

\author[M. Lang, K. Holley-Bockelmann, T. Bogdanovi\'c, P. Amaro-Seoane, A. Sesana,\& M. Sinha]
{M. Lang$^{1}$, K. Holley-Bockelmann$^{1,\,2}$, T.
Bogdanovi\'c$^{3,\,4}$, P. Amaro-Seoane$^{5,\,6}$, A. Sesana$^{5}$ \& M. Sinha$^{1}$\\
$^{1}$Department of Physics and Astronomy, Vanderbilt
University, Nashville, TN email: {\tt meagan.lang,\,manodeep.sinha@vanderbilt.edu}\\
$^{2}$Fisk University, Department
of Physics, Nashville, TN email:{\tt k.holley@vanderbilt.edu}\\
$^{3}$Department of Astronomy, University of Maryland,
College Park, MD 20742, e-mail: {\tt tamarab@astro.umd.edu}\\
$^{4}$Einstein Postdoctoral Fellow\\
$^{5}$Max Planck Institut f\"ur Gravitationsphysik
(Albert-Einstein-Institut), D-14476 Potsdam, Germany, email: {\tt pau.amaro-seoane,\,alberto.sesana@aei.mpg.de}\\
$^{6}$Institut de Ci{\`e}ncies de l'Espai (CSIC-IEEC), Campus UAB,
Torre C-5, parells, $2^{\rm na}$ planta, ES-08193, Bellaterra,
Barcelona, Spain}

\begin{document}
\date{}

\pagerange{\pageref{firstpage}--\pageref{lastpage}} \pubyear{2012}

\maketitle

\label{firstpage}

\begin{abstract}
Observations of the Galactic centre (GC) have accumulated a multitude
of ``forensic'' evidence indicating that several million years ago the
centre of the Milky Way galaxy was teeming with star formation and
accretion-powered activity -- this paints a rather different picture
from the GC as we understand it today. We examine a possibility that
this epoch of activity could have been triggered by the infall of a
satellite galaxy into the Milky-Way which began at the redshift of
$z=\newredshift$ and ended few million years ago with a merger of the
Galactic supermassive black hole with an intermediate mass black hole
brought in by the inspiralling satellite.
\end{abstract}


\section{Introduction}\label{S_intro}

There is mounting observational evidence that the epoch that ended
several million years ago was marked by an unusual level of activity
in the Galactic centre (GC). This is remarkable given that at the
current epoch, the GC is best characterized by the quiescent and
underluminous nature of ${\rm Sgr~A^*}$\citep{genzel10}. The picture of the 
GC as a once
powerful nucleus has begun to emerge from circumstantial observational
evidence, most recently strengthened by a discovery of the ``Fermi
bubbles'', a pair of giant gamma-ray emitting bubbles that extend
nearly 10~kpc north and south of the GC~\citep{dobler10,su10}.  
Although there are alternative steady state models for forming
the bubbles \citep{crocker11}, the well defined shock fronts at 
their edges suggest an abrupt origin. Current explanations include a 
past accretion event onto the supermassive black hole~\citep[SMBH;][]{su10,zubovas11}, 
AGN jets~\citep{guo11}, a nuclear starburst~\citep{su10}, and a sequence of
star capture events in the last $\sim$10~Myr~\citep{cheng11}.  The
period of increased gamma-ray activity is consistent with the finding
that until several hundred years ago ${\rm Sgr~A^*}$ was orders of
magnitude more X-ray luminous than it is today, as indicated by the
echo in the fluorescent Fe~K line emission detected in the direction
of the molecular clouds in the vicinity of ${\rm Sgr~A^*}$
\citep{inui09,ponti10,terrier10}. Although we cannot be certain that
the current quiescence of the GC is unusual, it appears clear that the
GC experienced an active phase as recently as a few hundred years ago.

The GC is also a hotbed of star formation containing the three most
massive young star clusters in the Galaxy: the Central cluster, the
Arches cluster, and the Quintuplet cluster \citep[see][for a
review]{figer08}. The three clusters are similar in many
respects. Each cluster contains $\sim 10^4 \Msun$ in stars and has
central stellar mass density that exceeds those measured in most
globular clusters. While our current understanding of massive star and
star cluster formation is incomplete, it is plausible that these
clusters are all characterized by the star formation event within the
past 2$-$7~Myr that resulted in the formation of more massive stars (above
$100 \Msun$) than anywhere else in the Galaxy
\citep{krabbe95,paumard06}. It is possible that the three clusters
have a common origin and that they have formed as a consequence of a
single event that triggered the flow of the copious amounts of gas
into the central $\sim 50$~pc in the Galaxy \citep[though see][for a
scenario in which the Arches cluster forms at the intersection of
X1/X2 gas orbits in the inner Galaxy]{StolteEtAl08}.

On even smaller scales, the existence of massive and young stars in
the Central cluster, well within the central parsec, is especially
puzzling
given their close proximity to the central SMBH.
Among those scenarios proposed are: in-situ formation \citep{bonnell08,mapelli12},
inspiral and consequent disruption of a dense stellar cluster with a central 
intermediate mass black hole (IMBH) \citep{merritt09}, and 
binary disruption by massive perturbers \citep{perets10}.
A clue in favor of the in-situ
formation is that most 
O and Wolf-Rayet type stars at the galactic center
seem to inhabit one or
more disc-like structures, pointing to their birth in a dense
accretion disc~\citep{bartko10,bonnell08,mapelli12}.
Star formation in a gaseous disc also
provides a natural explanation for the cuspy distribution of the young
stars. In order for starforming clumps to withstand tidal forces in
the inner parsec of the GC their densities need to be in excess of
$10^{11}\,{\rm cm^{-3}}$, at least five orders of magnitude higher
than the average density of molecular clouds in the
GC~\citep{figer00}. Such densities can only be achieved through highly
compressive events~\citep{figer08} and it is plausible that both the
inflow of large amounts of gas into the GC and its shocking and
compression have been caused by a common culprit. Both phenomena are
found to arise as consequences of galactic
mergers~\citep{noguchi88,bh91,bh92,bh96,mh96,hq10} making this a
possibility worth examining.

Further evidence that the MW has recently survived a dramatic event
comes from the distribution of late-type stars in the GC.
While the early-type stellar distribution appears to be
cuspy~\citep{genzel03,paumard06,buchholz09,do09,bartko10}, there seems
to be a distinct lack of late-type stars. This evidence is 
based on number counts of spectroscopically identified 
late-type stars brighter than magnitude $K=15.5$ within the 
sub-parsec region about Sgr~A$^*$. The best fits of the inner density 
profile for the late-type stellar population seem to
favor power-laws with slopes of $\gamma < 1$ and even allow the
possibility of a core with $\gamma < 0$, with the stellar density
decreasing toward the centre~\citep{buchholz09,do09,bartko10}. At this
stage, the evidence for a deficit of late-type stars is compelling,
however there are still significant uncertainties in the density
profile: the population of stars on which this inference has
been made are luminous late-type giants that comprise only a small
fraction of the underlying stellar density of the late-type
population. Regardless of the precise slope, however, the distribution
of the late-type population is contrasted by the steeply rising 
density distribution of early-type stars.

Possible mechanisms that could create a core in the distribution of
late-type stars have been discussed by~\citet{merritt10} and include
1) stellar collisions that strip red giants of their envelopes such that 
they are under-luminous, 2) destruction of stars on orbits that pass close to the 
SMBH in a triaxial nucleus, 3) inhibited star formation near the SMBH at the time
when the late-type population was formed, and 4) ejection of stars by a massive
black hole binary. In light of the other evidence that points to a
discrete event in the recent history of the GC, it is interesting to
revisit the latter mechanism.

In giant elliptical galaxies, the existence of cores is often
attributed to ejection of stars by an inspiralling binary
SMBH~\citep{merritt01,faber97,ferrarese06,mm01} and possibly due to gravitational wave
recoil after binary coalescence~\citep{bk04,gm08}. The prediction of
these models is that the central stellar mass deficit (traced by the
stellar light) is proportional to the mass of the central black hole,
$M_{\rm def}\propto M_{\bullet}$ \citep{graham04,hh10}. This correlation is interesting in
view of the observed dichotomy between ellipticals with cores and
those with the extra central light: core light deficit was found to
correlate closely with $M_{\bullet}$ and stellar velocity dispersion
$\sigma$, in agreement with the theoretical
predictions~\citep{mm01,gm08}, however, the extra light does
not~\citep{kb09}. An explanation of these phenomena offered
by~\citet{kb09} is that the extra light ellipticals were made in wet
mergers with starbursts, where stars formed from gas leftover after
the merger, while core ellipticals were created in dry mergers. In
galaxies with excess light, the newly formed population of stars fills
the core left in the distribution of the older population to form a
steep cusp, thus giving rise to characteristic differences in the two
stellar populations that may be mirrored in the MW GC.

Because studies of the light excess and deficit in elliptical galaxies
focus on major mergers, the scenario seems less relevant for a disc-dominated
system like the Milky Way, which may have never experienced a major
merger~\citep{gilmore02}. A minor merger of the SMBH with an 
IMBH 
however cannot be ruled out. The
presence of an IMBH in the Galactic centre has been previously
considered as a possible vehicle for delivery of young stars into the
GC~\citep{hm03}, a mechanism for creation of hypervelocity
stars~\citep{baum06}, and for the growth of the SMBH~\citep{pz06}. Indeed, the
possibility that an IMBH with mass $\lesssim 10^4 \Msun$ is still
lurking in the inner parsec of the GC cannot currently be totally excluded based
on observations~\citep{hm03,genzel10,rb04,gm09,ggm10}.

In light of the new observational evidence, which supports the notion
that few to 10~Myr ago was a special period in the life of ${\rm
Sgr~A^*}$, as indicated by the relatively recent episode of star
formation and increased energy output, we revisit the possibility that
a minor merger could have triggered this epoch of enhanced
activity. We suggest that the cumulative observational evidence favors
the minor merger hypothesis relative to the scenarios that propose a
steady state evolution or passive relaxation of the GC region. We present a
theoretical scenario for one such minor merger in \S~2 and
discuss the implications in \S~3.

\section[]{Milky Way -- Satellite Merger Scenario}\label{S_methods}

Here we examine the viability of the following scenario: 
 at high redshift, 
a primordial satellite galaxy with a central IMBH begins to
merge with a young Milky Way. As the satellite sinks toward the GC
under the influence of dynamical friction it is tidally stripped and
its orbit gradually decays toward the Milky Way disc
plane~\citep{qg86,callegari11}. The satellite perturbs previously
stable gas clouds in the inner Milky Way disc, driving gas
inflow~\citep{noguchi88,bh91,bh96,hq10} and compressing the gas to
densities exceeding those necessary for massive star formation near
the GC~\citep{mh96,hq10}. The satellite galaxy is expected to be
largely disrupted by the time it reaches the GC, leaving the IMBH
spiraling in a dense gaseous and stellar environment.  In the context
of this scenario we hypothesize that the IMBH reached the central
parsec on the order of $\sim$10 Myr ago. A fraction of perturbed gas
that did not form stars accretes onto the Milky Way's
SMBH~\citep{hq10}, injecting massive amounts of energy into the
surrounding medium and giving rise to the Fermi
bubbles~\citep{su10,zubovas11}.  Once gravitationally bound, the
IMBH-SMBH binary orbit tightens via three-body interactions with
surrounding stellar background, scouring the old stellar population to
form a central core~\citep{merritt10}.  Finally, the binary coalesces
after emitting copious gravitational radiation~\citep{pm63}.

In the context of this hypothetical scenario we use the new GC
observations to constrain the initial masses of the satellite and
Milky Way galaxies ($M_{\rm sat}$ and $M_{\rm MW}$), the mass deficit
in the late-type stellar population ($M_{\rm def}$), the IMBH mass ($M_{\rm
IMBH}$), as well as the amount of gas inflow into the GC triggered by
the inspiral of the satellite galaxy. The properties of the satellite bound to reach the inner disc of the Milky Way and deliver its IMBH to the GC must satisfy several criteria: 1. it should be light enough not to disrupt the Galactic disc, 2. it should be sufficiently massive in order for the dynamical friction to operate efficiently and deliver it to the GC within a Hubble time, and 3. its potential well should be sufficiently deep to sustain tidal stripping by the Milky Way. We therefore focus on constraining the most plausible scenario given the current understanding of the processes involved. 

Our approach is, out of necessity, semi-analytical in nature. While advanced 
cosmological nbody simulations are capable of modeling the accretion of a
low-mass satellite galaxy onto cosmologically growing Milky Way halo, there
are a number of physical processes important to our model that these 
simulations cannot capture. For example, we will capture the effect of
the Milky Way disc, bulge and SMBH, and will account for the stabilizing 
effect of the IMBH within the satellite. In our model, we also include
the critical effects of 3-body scattering and gravitational wave emission, 
both of which are beyond the reach of a cosmological nbody simulation.

\subsection{Properties of the Progenitor Milky Way}\label{SS_gal}
Beginning the merger at high redshift is advantageous in three
respects.  First, at this early epoch, it is reasonable to assume that
the proto-Milky Way was surrounded by primordial satellite galaxies
capable of housing a central seed black
hole~\cite[e.g.,][]{rg05,gk06,wa08,micic11}.  Second, at this stage in
its growth, the Milky Way would have been smaller, less massive, and
more gas-rich than it is today, thus decreasing the time required for
the satellite to sink to the galactic centre via dynamical friction.
Finally, the orbits of infalling satellites are more radial at high
redshift, which further shortens the merger time-scale~\citep{wetzel11}.
It should be noted that, while the remainder of this work posits that the satellite is accreted at 
redshift 8, this is by no means a unique solution.

To determine the properties of the Milky Way at this epoch, we assume
that it grows according to the exponential halo model from
\citet{mcbride09}:
\begin{equation}
M(z) = M_{z=0}(1+z)^{\beta}{\rm exp}\left(-\ln2\frac{z}{z_{f}}\right), \label{eqn:halogrowth}
\end{equation}
\noindent where $M_{z=0}$ is the current halo mass and $z_f$ is the
formation redshift, defined as the redshift at which the halo has
grown to half its current mass. Adopting the properties for the Milky
Way at $z_{f}=1$ as $M_{z=0}=2\times10^{12} \Msun$ and
$\beta=0.25$, the Milky Way's mass at $z=\newredshift$ 
can be estimated to be
$M_{\rm MW} = \newmwminit \Msun$. Studies of cosmological N-body simulations have found that at the redshift considered, the concentration of dark matter haloes is very weakly dependent on mass \citep{zhao03,gao2008,klypin11}. Following the methods outlined by \citet{prada11}, we find that at $z=8$ a halo of this mass will have a concentration of $c(z=8)\sim6$. The halo virial radius in a LCDM cosmology is defined as the radius where the mean enclosed density is 96 times the critical density 
of the universe, $\rho_{\rm crit}$. With the definition of  $\rho_{\rm crit}$:

\begin{equation}
{\rho_{\rm crit}} = {{{3 H_0^2}\over {8 \pi G }} \left[\Omega_{\Lambda}+ (1+z)^3 \, \Omega_{m} \right]},\label{eqn:rhocrit}
\end	{equation}

\noindent where $\Omega_{\Lambda}=0.73$ is the fraction of energy density in the universe in vacuum energy, while $\Omega_{m}=0.27$ is the fraction of energy density in the universe in matter, and z is the redshift. We find that the progenitor MW halo has a virial radius of $\sim 6$ kpc. This implies the progenitor Milky Way halo will have a density at 10 pc of 
$\newmwrho \Msun/$pc$^{3} \sim 10^6\rho_{\rm crit}$. 

It is important to note that while Eqn. \ref{eqn:halogrowth} assumes a single, smoothly growing
Milky Way halo, at these high-redshifts, mergers with other massive haloes are
very common, and the halo grows in a step-wise fashion.~\citep{diemand07}. 
Indeed, the entire picture of a single, virialized progenitor Milky Way halo is not strictly correct, and the `Milky Way' at this redshift 
is more likely a set of several haloes, many of which have not yet decoupled from the Hubble flow to allow
turnaround and collapse into a single virialized structure. Consequently, our assumption of a virialized NFW 
halo at the accretion redshift ($z=\newredshift$) must be recognized 
as an approximation made due to the limits of a semi-analytic approach.

\subsection{Finding the Culprit Satellite}\label{SS_orb}

Broadly, we identify possible culprit satellites by integrating the
orbits of infalling haloes within an analytic, but evolving Milky Way potential. As both the satellite and Milky
Way evolve, we search for the satellites that reach the Inner Lindblad Resonance (ILR) at 150 pc roughly
10 Myr ago after losing over 95\% of its initial orbital angular momentum. Of the satellites that survive 
until they plunge through the  ILR, we preferentially select those that retain enough mass to perturb the gas there. 
The culprit satellite is characterized by the mass, radius and concentration, as well as the energy, angular momentum, infall radius and merger redshift of the orbit.
We elaborate on the procedure below.

We adopt a merger redshift of  $\sim\newredshift$. In order to deliver the IMBH to the GC a mere 2$-$7 Myr ago, the
proposed merger redshift implies 
that the satellite orbit
decayed over a time-scale of about \newdecaytime{ }Gyr.  
At such a high redshift,
the IMBH and satellite had very little time to evolve before being
accreted by the Milky Way, making the pair a ``fossil'' of the dark ages
before reionization \citep{rg05,gk06}. 

We rely on cosmological N-body simulations to constrain the
initial conditions of the orbit. These inform us that at the present
epoch, satellites are preferentially accreted on very eccentric
orbits, with a distribution peak at about $e=0.85$
\citep{benson05,wang05,zentner05,khochfar06,ghigna98,tormen97}. At
higher redshifts the satellite orbits are characterized by even higher
eccentricities, albeit, in both cases the distribution peaks are
broad. Seemingly independent of redshift, a typical satellite is
accreted at the virial radius with a total velocity, $|\vec{v}_{\rm sat}|=1.15 v_{\rm vir}$ ($v_{\rm
vir}$ is the circular velocity at the virial radius of the primary
galaxy) that marks it as barely bound
\citep{benson05,wetzel11}. Motivated by these results,
we select an orbit that has $|\vec{v}_{\rm sat}|=1.15 v_{\rm
vir}$ at the virial radius of the primary and an eccentricity of
$\newecc$, consistent with expectations for the eccentricity 
distribution peak at $z=\newredshift$ \citep{wetzel11}.

Starting with the above total velocity and eccentricity, we calculate 
the orbital decay for a range of satellite masses placed at the 
virial radius of the primary. For a given initial position at the virial radius, 
the azimuthal and radial components of the satellite's initial velocity within the orbital 
plane are calculated in terms of the eccentricity ($e$) and total velocity ($|\vec{v}_{\rm sat}|$) as:
\begin{equation}
v_{\phi} = \frac{v_{vir}}{|\vec{v}_{\rm sat}|}\sqrt{\frac{GM_{\rm MW}}{r_{\rm vir}}(1-e^2)} \mbox{ and } \\
v_{r} = \sqrt{|\vec{v}_{\rm sat}|^{2}-v_{\phi}^{2}}.
\end{equation}

We adopt an analytic model of the Milky Way that includes a
central SMBH, Miyamoto-Nagai thin disc~\citep{mn75}, a spherical
Hernquist bulge~\citep{hernquist90}, and an NFW halo~\citep{nfw97}.
To mimic a young Milky Way, we  use  Eqn. \ref{eqn:halogrowth} to set the halo mass. We set the virial radius using the mass and the critical density at the
starting redshift in Eqn. \ref{eqn:rhocrit},
and we initialize the concentration using \citet{prada11}. We assume that the mass and size of the baryonic components change in the same 
way as the halo does; this is not true in detail, but allows us to convert the known present-day Milky Way parameters to the starting
redshift.  Our current Milky Way mass model is similar to analytic models best-fit to rotation curve data~\citep[e.g.][]{wid05, den98} $z=0$ disc mass is $5\times10^{10} \Msun$, the disc scale length is $3$ kpc and, and the disc scale height is $300$ pc.For the bulge, we set a current epoch bulge mass of $8\times10^{9} \Msun$ and scale length of $0.7$ kpc.

We integrate the orbits using a fourth-order Runge-Kutta method 
to step the satellite's position and velocity forward in time. At each timestep, we adjust the
analytic Milky Way model using the method described above. We calculate the acceleration 
of the satellite due to this evolving analytic potential, and we  include  Chandrasekhar dynamical friction~\citep{chandra43}, as well as
mass loss from the satellite due to tidal stripping and disc shocks.
The acceleration due to dynamical friction is calculated in the uniform
 density limit as
\begin{eqnarray}
\left(\frac{d\vec{v}_{\rm sat}}{dt}\right)_{\rm fric} &=& -\frac{4\pi\ln
\Lambda G^{2}M_{\rm sat}\rho_{\rm MW}}{|\vec{v}_{\rm sat}|^{3}}\times
\cdots\nonumber\\
&&\times\left[{\rm erf}(\chi)-\frac{2\chi}{\sqrt{\pi}}e^{-\chi^2}\right]
\vec{v}_{\rm sat}, \label{eqn:accfric}
\end{eqnarray}
where $M_{\rm sat}$ is the mass of the satellite, 
$\rho_{\rm MW}$ is the density of the Milky Way at the satellite's 
position, 
$\ln\Lambda = \ln\left[1+ (M_{\rm MW}/M_{\rm sat})^2\right]$ 
is the Coulomb logarithm, 
$\chi=|\vec{v}_{\rm sat}|/\sqrt{2}\sigma$, and 
$\sigma=\sqrt{GM_{\rm MW}/2R_{\rm MW}}$ is the average velocity 
dispersion of the Milky Way halo.

At each step in the orbit, we calculate the local density of the Milky Way and we tidally strip the satellite to the Roche radius, where the density 
of the satellite is equal to the Milky Way background. We also model mass loss from disc shocking by removing 
\begin{equation}
\Delta M_{\rm shock} = \frac{5}{3}\frac{4}{GM_{\rm sat}v_{\rm sat,z}^{2}}\left(\frac{dv_{\rm sat,z}}{dt}\right)_{\rm disc}^{2} \label{eqn:shock}
\end	{equation}
from the satellite's mass each time it passes through the Milky Way disc \citep{gnedin97}.
  We neglect the stellar component of
the satellite, since the baryon content of such low mass satellites is
relatively uncertain, but likely to be very small
\citep{gnedin00,sg07,ricotti08}.

We find that the most likely culprit
is a satellite with a mass of $M_{\rm sat} \approx \newsatminit \Msun$. Modeling 
the satellite dark matter profile as an NFW halo,
its corresponding concentration parameter at this redshift is about
\newsatconcen~\citep{prada11}, making the satellite's central
density within the inner 10 pc $\sim\newsatrho \Msun/$pc$^{3} \sim 4 \times 10^5 \,\rho_{crit}$ or $\sim 2\times10^4$ times the Milky Way's density
at the virial radius. Including an IMBH in our satellite model would deepen its central potential 
and could aid in delivering the satellite core to the centre of the MW intact, although we did not 
include this effect in our calculations.

By the time the satellite has reached the
inner 100 pc, it will have lost most of its mass, with 
$\sim \newsatmfinal \Msun$ remaining.  Without direct hydrodynamic simulations,
it is difficult to say how much damage this IMBH-embedded satellite core could do to
the gas-rich inner Milky Way.  In general, we expect the satellite to perturb the gas in the
galactic centre, torquing it and transporting angular momentum through narrow resonances~\citep{gold79}; the classical
rate of gas inflow from this process is proportional to the strength of the perturbation squared.
However, when the system has a significant asymmetric perturbation, the orbits begin to cross one another and gas piles up in shocks~\citep{papa77}. In this case,
 the radial inflow rate of gas from a global perturbation is linearly proportional to the strength of perturbation~\citep{hq11}, and numerical simulations find
 the shocks induced by even a few $\%$ perturbation can destabilize
the gas and drive gas inflow \citep{bh91,bh96,mh96,hq10}. 
To estimate the perturbation a $\sim \newsatmfinal \Msun$ 
satellite core could exert on the gas accumulated in a ring at the Inner Lindblad 
Resonance (ILR) of the Milky Way, we refer to ~\citet{vw00}, which explores the perturbation strength
induced by galaxy flyby encounters.   Using linear perturbation theory, ~\citet{vw00} find that a flyby with a mass ratio of 10 and a pericentre at the half-mass radius will induce
a strong perturbation in the density of the primary galaxy of order unity.  Since the mass ratio of the
 inner Milky Way ($\sim10^{8} \Msun$) to the satellite remnant
is 1000 \citep{lindqvist92}, we expect a perturbation of the order $|a|\sim0.01$ in the surface density.
The linear relationship between gas inflow and perturbation amplitude derived by \citet{hq11},
\begin{equation}
\frac{dM_{\rm gas}}{dt}=|a|\Sigma_{\rm gas}R^{2}\Omega,
\end{equation}
can then be used to gauge the expected amount of gas inflow. 
Setting the perturbation amplitude to $|a|\sim0.01$, 
the radius to $R=150$ pc (ILR),
the rotation frequency to $\Omega(R)=v_{circ}(R)/R=0.62$ Myr$^{-1}$ \citep{stark04}, and
the gas surface density to $\Sigma_{\rm gas}=500\Msun/$pc$^{2}$ based on observations of other
barred galaxies \citep{jogee05} and of the molecular ring in the MW GC \citep{molinari11}, 
yields a gas inflow rate of $\sim7\times10^4 \Msun/$Myr. Assuming this inflow rate over $\sim10$ Myr,
 we find that this satellite should be able to drive a 
net inflow of $\sim10^{6}\Msun$ of gas from the ILR.

\subsection{Late-Type Stellar Mass Deficit and IMBH Mass}\label{SS_mdef}
If the core in the distribution of late-type stars at the GC
\citep{buchholz09,do09} was scoured out by an IMBH-SMBH binary~\citep{preto11,gm12}, 
the
amount of stellar mass missing from the GC can be used to constrain
the mass ratio of the black hole binary~\citep{mm01,gm08,merritt06}. 
To determine this mass deficit we compare the stellar distribution
inferred from observations with that expected for a dynamically
relaxed system without a core. In terms of the number density of the
late-type stellar population, the core can be represented by a broken
power law
\begin{equation}
\label{dens_fin}
n_{f}(r) = n_{0}\left(\frac{r}{r_{0}}\right)^{-\gamma_{i}}
\left[1+\left(\frac{r}{r_{0}}\right)^{\alpha}\right]^{(\gamma_{i}-\gamma)/\alpha},
\end{equation}
\noindent with $n_{0}=0.21$~pc$^{-3}$, $r_{0}=0.21$~pc, $\gamma=1.8$,
$\gamma_{i}=-1.0$, and $\alpha=4$ \citep{merritt10}. We adopt this
description in our analysis but note that in presence of strong mass
segregation the slopes can be steeper
\citep{AlexanderHopman09,PretoAmaroSeoane10,Amaro-SeoanePreto11}. The
observed distribution of stars outside of the $0.21$~pc core radius is
consistent with the Bahcall-Wolf profile \citep[$\propto
r^{-1.75}$,][]{bahcall76} of a relaxed system as it would have existed
prior to scattering by the IMBH-SMBH binary.  We model the initial
stellar cusp by extending the $r^{-1.8}$ profile to smaller radii:
\begin{equation}
\label{dens_in}
n_{i}(r) = n_{0}\left(\frac{r}{r_{0}}\right)^{-\gamma}.
\end{equation}
Assuming that the mass density profiles {\it before} and {\it after} the creation
of the core are proportional to equations (\ref{dens_in}) and
(\ref{dens_fin}) respectively, we calculate the mass deficit as the
integrated difference between the initial and final (observed)
profiles. We normalize the profile given by Eqn. \ref{dens_fin} such that 
integrating it over the inner parsec yields 
$1.0\pm0.5\times10^{6} \Msun$, the mass determined 
by \citet{schodel09}, and obtain $M_{\rm def} \approx 2\times10^{5} \Msun$. 

It should be understood that this mass deficit can only be treated as an estimate.
Although this calculation assumes the best fit core radius
($r_{0}$) and inner slope ($\gamma_{i}$) from \citet{merritt10}, the fit was not excellent 
($\tilde{\chi}^2>17$) and the estimated mass deficit is highly dependent on 
these parameters. In addition, this calculation assumes that core size 
(and therefore the mass deficit) has not 
changed significantly over time. This is consistent with a 
core scoured recently enough ($\sim10$ Myr) that relaxation has not yet had 
enough time to fill in the core \citep[$\sim10$ Gyr;][]{merritt10}. 
However, it is also possible that the core is an evolved system; this implies a larger core, 
more massive IMBH, and more dramatic scouring event in the distant past.
In this case, the creation of the core would have been unrelated to the creation of 
the young GC stars or Fermi bubbles.

N-body merger simulations studying the relationship between the ratio
of total stellar mass ejected to binary mass, $M_{\rm def}/(M_1+M_2)$,
and binary mass ratio, $q = M_1/M_2$, have not yet been carried out
for the mass deficit calculated here. In order to relate the two we use a
semi-analytic formalism describing the interaction of massive black
hole binaries with their stellar environment \citep{sesana08} to place
the upper and lower limits on the mass of the IMBH based on $M_{\rm
def}$ inferred from observations.

It has been shown by numerical simulations \citep{baum06,matsu07} and
semianalytic models \citep{sesana08}, that an IMBH inspiralling in a
stellar cusp surrounding a central SMBH starts to efficiently eject
stars at a separation $a_0$, where the stellar mass enclosed in the
IMBH orbit is of the order of $2M_2$. The ejection of bound stars
causes an IMBH orbital decay of a factor of $\approx 10$, excavating a
core of radius $r_0\approx 2a_0$ in the central stellar cusp,
resulting in a mass deficit about $3M_2$ \citep[see][for
details]{sesana08}. Such orbital decay is in general insufficient to
bring the IMBH in the efficient gravitational wave (GW) emission
regime, unless its eccentricity grows to $>0.9$ during the shrinking
process. It is also the case in this picture that the mass of the
inspiralling IMBH inferred for a given mass deficit strongly depends
on the eccentricity evolution of its orbit. In what follows, we
consider both the high and low orbital eccentricity scenario and use
them to place a bound on the plausible range of IMBH masses.

If the eccentricity grows efficiently, the IMBH depletes the central
cusp, forms a core of a size $\approx 2a_0$, and merges due to GW
emission on a time scale of only $1-10$~Myr~\citep{sesana08}. For a
stellar distribution described by an isothermal sphere outside of the
radius of influence of the SMBH, $a_0 = 2q^{4/5}$pc. Adopting the core
radius of $r_0=2a_0=0.21$ pc, we find $q=0.02$, and an upper limit on
the mass of the IMBH, $M_2 = 8\times10^4 \Msun$. In this case, the
mass evacuated from the stellar cusp by the IMBH is of the order of
$3M_2$~\citep{sesana08}, i.e., $\approx2.5\times 10^5$M$_\odot$,
consistent with the stellar mass deficit measurement in the GC.

Alternatively, if the IMBH eccentricity does not grow significantly
during the bound cusp erosion, further scattering of stars
replenishing the binary loss cone is needed in order to evolve from
separation of $a_0$ to the GW regime. Therefore, a circular orbit
regime can be used to establish a lower limit on the mass of the IMBH,
for a given mass deficit indicated by observations. We
assume that in this case both $r_0$ and $M_{\rm def}$ created in the
cusp erosion phase are small (we justify this assumption below). In
this scenario, the final $r_0$ and $M_{\rm def}$ are reached as a
consequence of the diffusion of the stars from the edge of the small
core into the loss cone of the binary. The ejections of each star
carry away an energy of the order $(3/2)G\mu/a$~\citep{quinlan96},
where $\mu=M_1\,M_2/M$. We compute $M_{\rm def}$  by imposing:
\begin{equation}
\frac{3}{2}\frac{G\mu}{a}dM_{\rm def}=\frac{GM_1M_2}{2}d\frac{1}{a}
\end{equation} 
to get
\begin{equation}
M_{\rm def}=\frac{M_1+M_2}{3}{\rm ln}\frac{a_i}{a_f},
\end{equation}
where $a_{\rm i}$ is the hardening radius of the binary (radius at
which the scattering of unbound stars becomes effective) and $a_{\rm
f}$ is the separation at which the GW emission becomes efficient.
Using equations (19) and (20) in~\cite{sesana10} to express $a_{\rm
i}$ and $a_{\rm f}$, it follows that,
\begin{equation}
M_{\rm def}=\frac{M_1+M_2}{3}\,{\rm ln}\frac{500\,q^{4/5}}{F(e)^{1/5}},
\end{equation}
where $F(e)=(1-e^2)^{-7/2}(1+73/24\,e^2+37/96\,e^4)$. Assuming for the purpose
of this estimate that the binary remains circular throughout its evolution and
imposing $M_{\rm def}=2\times 10^5 \Msun$, we find $q=5\times10^{-4}$ and a
lower limit on the mass of the IMBH is $M_2=2\times10^3 \Msun$ \footnote{Both
numerical simulations and semi-analytic models however suggest that the
eccentricity in the cusp erosion phase grows to $>0.9$, in which case
$F(e)>1000$ and $q>5\times 10^{-3}$, i.e., $M_2>2\times10^4 \Msun$.}.  An
IMBH of such mass, would excavate a core of $\approx 0.01$ pc, causing a mass
deficit of $\sim 3M_2=6\times10^3 \Msun$ in the bound scattering phase and thus,
justifying our earlier assumption that the diffusion of stars into the loss cone is the primary 
process that shapes the properties of the core in this case.  Note that in the circular orbit scenario
the time scale for the inspiral of the IMBH towards the GW regime is determined
by the unknown rate of diffusion of the stars into the loss cone of the binary.
Hence, depending on the time scale of relaxation processes this process could
in principle lead to the IMBH-SMBH binary ``hangup'', i.e., a long lived ($>
1$~Gyr) binary configuration at separation $<r_0$ -- tantamount to the
classical ``final parsec'' problem~\citep{BegelmanEtAl80}. 
It is however possible that the binary will not stall in our specific case.
The galactic centre in this phase will be described by a
strongly perturbed, non-axisymmetric potential which allows stars to
scatter into the loss cone efficiently
\citep{merritt04,berzcik06,perets08,khan11}.  Moreover, the orbit will occur
in a relatively gas-rich environment, which can further aid the decay
of the binary \citep{escala05,dotti07,cuadra09}. Finally, any extra
stars brought in by the satellite would help the binary decay~\citep[see][]{Miller02}.  
Even under the assumption of a circular orbit, 
an efficient coalescence can occur on a time-scale of 10~Myr.

This analysis suggests that the observed mass deficit and core size
are consistent with the IMBH mass in the range $2\times
10^3 \Msun <M_2<8\times 10^4 \Msun$, 
whereas the efficient eccentricity growth found in N-body
simulations and semi-analytic models favor
$M_2\gtrsim 10^4 \Msun$. Within this range, the time scale for the
IMBH to create a core and merge with the SMBH can be as short as few
Myr. On the other hand, a possibility that an IMBH may be still be
lurking in the GC is not completely ruled out. We discuss the consequences of the
latter scenario in the context of the observational constraints on the
presence of a second black hole in the Galactic centre in
\S~\ref{S_discuss}.

It is useful to consider whether a satellite galaxy with an initial
mass of $M_{\rm sat}\sim \newsatminit \Msun$
can host a $\gtrsim 10^4 \Msun$ IMBH.  While there are no observational constraints for
galaxies or black holes of this mass range, there are three leading
theories for IMBH formation at high redshift: `direct collapse' of
metal-free, low angular momentum gas into a $10^3-10^6 \Msun$
black hole~\citep{loeb94,begelman08}, an unstable supermassive star
that collapses into a $10^2-10^5 \Msun $ black hole
\citep{colgate67,quinlan87, baumgarte99}, or a Population III star,
which would leave behind seed black holes of $\sim1-10^3 \Msun$
between redshift 30$-$12~\citep{madau01,bromm02,wa08,clark11}. 
Even if the IMBH in our satellite started as a low mass Pop III 
seed in a somewhat turbulent environment with a mass of 
$\sim5 \Msun$~\citep{clark11}, it is plausible that it would reach the IMBH mass proposed
here through a combination of gas accretion and black hole mergers~\citep{KHB10}. 
In such a satellite galaxy, it
would require less than one percent of the gas to accrete onto a low
mass seed to form the IMBH $\gtrsim 10^4 \Msun$. 

Note that the massive seeds produced in a direct collapse typically
favor more massive haloes than the one we have proposed as our
culprit. This is because metal-free gas collapses most efficiently in
haloes with $T_{\rm vir} > 10^4$ K, corresponding to $M_{\rm vir} >
10^8 \Msun [(1+z)/10]^{3/2}$ \citep{bromm03}. In the context of the
merger hypothesis choosing a slightly more massive satellite would
push the accretion redshift closer to the present day, and as long as
the resulting satellite merger is still a minor one, this does not
significantly affect the outcome of our scenario.

\subsection{Inflow of Gas and Gamma-ray Bubbles}\label{SS_bubble}
As noted in \S~\ref{SS_gal} the inspiral of a satellite galaxy can
cause the inflow of a significant amount of gas towards the centre of
the Galaxy~\citep{noguchi88,bh96,mh96,cox08}.  One fraction of this gas could have
given rise to the star formation in the Central, Arches, and
Quintuplet clusters, which marked the epoch between 2$-$7~Myr ago in
the central 50~pc of the Milky Way. All three clusters contain some of
the most massive stars in the Galaxy and have inferred masses of
$\sim10^4 \Msun$ \citep{figer08}. Assuming a ``standard'' star
formation efficiency of 10\%~\citep{ry99}, it follows that the amount
of gas necessary to produce the stellar population of the three
clusters is a ${\rm few} \times 10^5 \Msun$. Note that a sequence of
strongly compressional events during the satellite-Milky Way merger
could have given rise to a higher efficiency of star
formation~\citep{dimatteo07}, in which case the estimated mass of the
gas represents an upper limit.

In this merger scenario, the remainder of the perturbed gas that did
not form stars would be channeled towards the central parsec \citep{loose82}, and the
fraction that is accreted into the SMBH could drive the energetic
outburst of several Myr ago. The far-IR and millimeter observations
indicate that $\sim 10^4 \Msun$ of the molecular gas continues to
reside in the circumnuclear disc within the central $\sim1.5$~pc of
the Galaxy \citep[see][for review and references
therein]{genzel10}. The maximum amount of the remnant molecular gas
that has not been accreted onto the SMBH can also be estimated based
on its expected gravitational effect on the orbits of the stars
residing within the inner 0.5 pc. In this case, the requirement for
stability of the stellar disc over its lifetime of 6~Myr poses a
constraint on the mass of the molecular torus of $<10^6 \Msun$
\citep{subr09}.

On the other hand, the recent discovery of the two large gamma-ray
bubbles extending from the GC above and below the galactic plane are
compelling evidence of a relatively recent period of intense activity
in the now quiet GC. The gamma-ray bubbles exhibit several striking
properties: they are perpendicular and symmetric with respect to the
plane of the Galaxy, have nearly uniform gamma-ray brightness across
the bubbles, and well defined sharp edges \citep{dobler10,su10}. The
gamma-ray emission from the bubbles is characterized by the hard
energy spectrum and is most likely to originate from the inverse
Compton scattering of the interstellar radiation field on the cosmic
ray electrons -- the same population of electrons deemed responsible
for the diffuse synchrotron microwave radiation detected by the WMAP
\citep{finke04,df08}. The sharp edges of the Fermi bubbles are also
traced by the X-ray arcs discovered in the ROSAT maps
\citep{snowden97}, suggested to be the remnants of shock fronts
created by the expanding bubbles \citep{su10,guo11}.

The morphology, energetics, and emission properties of the Fermi
bubbles favor the explanation that bubbles were created in a strong
episode of energy injection in the GC in the last $\sim 10$~Myr that
followed an accretion event onto the SMBH \citep{su10}. Simulations by
\citet{guo11} indicate that the bubbles could have been formed by a
pair of bipolar jets that released a total energy of $1-8\times
10^{57}$~erg over the course of $\sim$0.1$-$0.5~Myr between 1 and
2~Myr ago. This explanation for the Fermi bubbles implies that $\sim
10^4 \Msun$ of material must have been accreted onto the SMBH at
nearly the Eddington rate, assuming the accretion efficiency of $10\%$
\citep{ss73,dl11}. Based on the range of models explored by
\citet{guo11} it is possible to estimate that the amount of mass
processed in such jets (i.e., the mass of the gas that fills the jet
cavities) is as small as $30 \Msun$ and as large as
$3\times10^5\Msun$.

This estimate, together with the gas that formed stars, the gas
accreted onto the SMBH and the gas processed by the jets allows us to
put a constraint on the total gas inflow into the central $\sim50$~pc
of the Galaxy of $\lesssim 10^6 \Msun$, consistent with the amount
expected from the perturbation analysis of the stability of the ILR
gas in the Milky Way.

\section[]{Discussion}\label{S_discuss}

\subsection{How rare are satellite merger events?}
We propose that the timeline began about \newdecaytime{} 
Gyr ago, when the proto
Milky Way accreted a small satellite dark matter halo at the time when
their haloes were physically closer and less massive. The satellite
orbit decayed slowly and only reached the GC a few million years ago,
after having been stripped of most its mass. The thinness of the Milky
Way disc has often been used as an argument against a recent minor
merger~\citep{qhf93,snt98,vw99}; however, the proposed satellite is so
minor, particularly by the time the orbit decays to 10 kpc, that the
thin disc could have survived unscathed
\citep{toth92,walker96,tb01,hopkins08,hopkins09}. 

Using the Extended Press-Schechter formalism (EPS)~\citep{bond91,bower91,lc93,pch08}, we can estimate the number of satellite accretion events a typical Milky Way
mass galaxy will undergo. We determined this on the basis of 100 realizations 
of an EPS merger tree that resulted in a base halo of $2\times 10^{12} \Msun$ at z=0.
In this calculation, we assumed WMAP5 parameters and summed over the haloes
in the $M_{\rm sat}=10^7-10^9 \Msun$ mass range that merged with the 
main halo from z=7 to z=0.We found a mean of 1745 such satellite accretion 
events, with a standard deviation of 425. 
However, about half of these accretions occur after $z=1$ --- 
and are unlikely to have made it to the GC by $z=0$. We confirmed that this number of satellite accretion events is
consistent with expectations from the cosmological simulations by
comparing to one of our N-body simulations of a 50 Mpc$^3$ volume described in ~\citet{shb11}.



%
\begin{figure*}
\includegraphics[height=7in,angle=270]{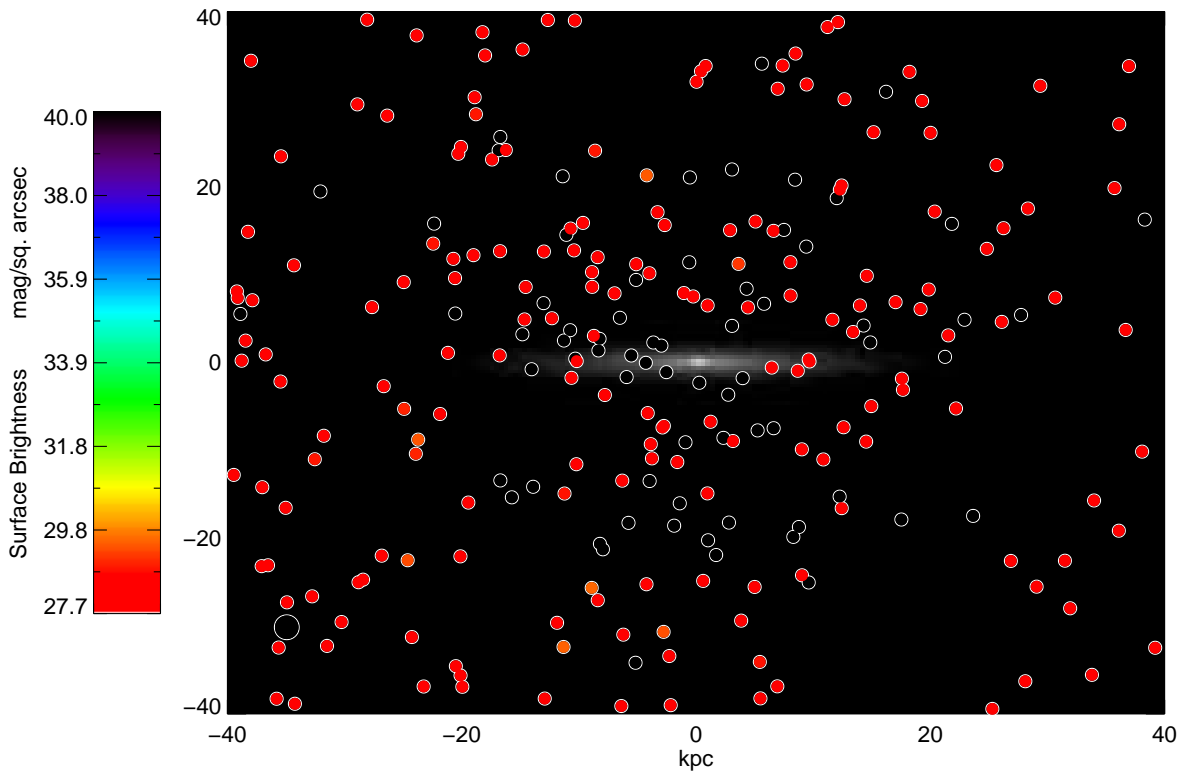}
\vspace{-0.6in}
\caption{Distribution of accreted low mass satellites at the present
day. The inner 40 kpc region of the Milky Way disc is shown in
greyscale with current accreted satellite positions overplotted. The
color maps to the surface brightness, and the relative size
corresponds to the tidal radius of the satellite. Note that the circle
size is not to scale and that none of these satellites would be
observable above the background. Also note that satellites that 
have merged with SMBH or completely disrupted are not plotted.}
\label{fig:conveyer}
\end{figure*}
Figure~\ref{fig:conveyer} illustrates one realization of the current
distribution of accreted satellites in the mass range $M_{\rm
sat}=10^7-10^9 \Msun$. We used the EPS technique described above to
define the number of accreted satellites in this mass range at each integer
step in redshift from z=7 to z=0. To define the orbit of each satellite 
as it is accreted, we randomly selected from the energy and
angular momentum distributions at each redshift using the expressions
$7-9$ in~\citep{wetzel11}.  As in section 2, we integrated orbits of the
satellites from the accretion epoch to the present day, scaling the
Milky Way mass and size to the redshift of
accretion using Eqn.~\ref{eqn:halogrowth}. Figure~\ref{fig:conveyer} 
shows the inner 40~kpc of the
current-day Milky Way; approximately 85\% of the accreted satellites
are at separations larger than 40~kpc, and only 5 reached the GC and merged with the SMBH. We estimate the surface
brightness of the satellites assuming that the baryons are confined to
a radius within the dark matter halo ten times smaller than the
satellite virial radius. We infer the initial star fraction from
\citet{rg05} and assume a total mass-to-light ratio of
 $\sim300$ \citep{strigari08} for the bound stars that remain 
after tidal stripping.

While it is very clear that not all of the small satellites can 
reach the
GC, what fraction does is a question of some subtlety. Galaxy merger
time-scales cited in the literature, particularly for the small mass
ratios considered here, span a wide range. The key to the
uncertainties is the treatment of dynamical friction: most
semi-analytic works, including this one, rely on the dynamical
friction formalism as described by Chandrasekhar (1943) but change the 
Coulomb logarithm to account for inhomogeneous or anisotropic systems~\citep{pen04,just05}, or to include mass loss~\citep{tb01,vw99}.
This approach has been shown to underestimate the decay time in pure
dark matter simulations \citep{colpi99,bk08}. On the other hand, the 
presence of gas can dramatically decrease the
orbital decay time of a satellite by efficiently dissipating 
its orbital energy throughout the system 
\citep{ostriker99,sanchez99} -- thus,
making the Chandrasekhar formula a significant overestimate.
Compounding the issue, linear perturbation theory and limited N-body
experiments indicate that resonant heating caused by orbits in the
satellite galaxy that are commensurate with the orbit of the satellite
about the GC can enhance mass loss and can change the angular
momentum of the orbit in non-trivial ways~\citep{weinberg97,choi09}. 
Although the Milky Way has likely accreted over a thousand of
these small satellites, it is uncertain how often they reached the
galactic centre. It is however plausible that the GC has experienced
a handful of these accretion events spread over its lifespan.
Despite the uncertainty that arises in the mass of our culprit satellite due
to the imprecise dynamical friction time-scale, the scenario itself
remains viable, because other constraints on satellite mass (satellite evaporation,
disc disruption) are flexible so long as the merger 
time-scale remains less than the age of the universe.

\subsection{Hypervelocity stars and stellar core}

While the properties of the newly formed stars and perturbed gas were
dictated by the accreted satellite, the IMBH was responsible for
carving out the old stellar population. As a gravitationally bound
binary IMBH-SMBH formed and decayed, it scoured out $2 \times 10^5
\Msun$ of the relaxed old and initially cuspy stellar population. Many
of these stars could have been ejected from the GC as hypervelocity
stars \citep[HVSs;][]{brown05,baum06}, though most may simply have received
enough energy to traverse the inner parsec. Simulations of IMBH-SMBH
binaries in stellar environments indicate that HVSs are created in a
short burst which lasts only a few Myr in case of a $\sim
10^4 \Msun$ IMBH \citep{baum06, sesana08}.  In the context of our
picture we predict that this event created $\sim 10^3$ hypervelocity
stars that, if they were ejected at about 1000~${\rm km\,s^{-1}}$
\citep{baum06}, ought to lie $\sim10$ kpc from the GC today. It is worth
noting that about a dozen of HVSs observed in the Galactic halo thus
far have travel times that span 60$-$240~Myr and appear to be
consistent with a continuous ejection model
\citep{brown08,brown09,tillich09,irrgang10} and not with the IMBH-SMBH
binary picture \citep{brown08,sesana08}. Along similar lines, the
spatial and velocity distribution of the current observed HVSs seem to be
inconsistent with a IMBH-SMBH slingshot origin \citep{sesanaetal07}.

The large size of the observed GC core, $r_{0}=0.21$~pc, could be seen
as a challenge to any scenario involving 3-body scattering, since
state of the art high resolution direct N-body simulations that
modeled the ejection of hypervelocity stars from a SMBH-IMBH binary in
the galactic centre never generated a core larger than 0.02 parsecs
\citep{baum06}. However, there are several effects that could conspire
to cause the simulated core size to be a lower limit. First, the mass
of the simulated SMBH in \citet{baum06} is $3\times 10^6 \Msun$, which
would eject fewer stars than somewhat more massive Milky Way
SMBH. Second, the density profile was sharply curtailed by a factor of
$(1+r^5)$ in order to minimize the number of stars far from the SMBH;
this makes the spatial distribution of stars in the simulated nuclear
star cluster more centrally peaked relative to that in the GC, which
can also result in a smaller core. In general, though, it is important
to note that the size of the scoured core is a property that
sensitively depends on the density, the eccentricity, and kinematic
structure of the GC or on assumptions in the model used to represent
it. 

\subsection{Has IMBH-SMBH binary merged?}

We now return to the question whether the IMBH-SMBH binary has already
merged or whether the IMBH could still be lurking in the GC. As
discussed in \S~\ref{SS_mdef}, the N-body and semi-analytic modeling
of the GC favor the evolutionary scenarios in which the inspiral and
coalescence of the SMBH with a $M_2\gtrsim 10^4 \Msun$ IMBH is
relatively efficient. Moreover, there is currently no empirical
evidence for a second black hole in the central parsec. In order to be
consistent with the observations, the IMBH present in the GC would
have to have a mass $\sim 10^3 - 10^{4.5}\Msun$ and be either very
close ($\leq 10^{-3}$ pc) or at $>0.1$ pc from the SMBH 
\citep{rb04,gm09,ggm10,genzel10}.
An IMBH in this mass
range that reaches a separation of $10^{-4}$ pc would merge with the
SMBH in less than 10~Myr due to the emission of GWs, thus
severely restricting the amount of parameter space where the IMBH
and SMBH can exist in a long lived binary configuration.
Nevertheless, given the uncertainties in the binary mass ratio,
eccentricity, and the structure of the initial stellar cusp, the
presence of an IMBH in the GC cannot be entirely ruled out at this
point.

If on the other hand, the IMBH and SMBH coalesced several million
years ago, one possible signature of this event could be a SMBH recoil
caused by the asymmetric emission of GWs \citep{per62,bek73}. Current
astrometric observations of the reflex motion of
the SMBH put strong constraints on the allowed recoil velocity; the
SMBH cannot have velocity with respect to the Central cluster larger
than $3.5$ km/s (within $1\sigma$ error), at the distance
of the GC \citep{yelda10}. Similarly,  \citet{rb04} constrain the peculiar
motion of Sgr~A* in the plane of the Galaxy to $-18\pm7$ km/s
and perpendicular to the Galactic plane to $-0.4\pm0.9$ km/s, 
where quoted uncertainties are $1\sigma$ errors. 
There is however a caveat with respect to the interpretation of the
SMBH reflex motion: if the reference frame in which the reflex motion
is measured is based on the nearby gas and stars bound to the SMBH,
the resulting relative velocity of the SMBH will be zero because in
this case, the stars and the gas move together with the SMBH as long
as their orbital velocity is higher than the that of the reflex
motion. The radio and near-infrared reference frames in \citet{yelda10} are 
defined based on the nearby stars orbiting around the SMBH and are 
thus a subject to this caveat. The measurement of \citet{rb04} is however 
carried out in the reference frame defined by the extragalactic radio 
sources and can be used to test the recoil hypothesis. 

For $10^4 \Msun$ IMBH the black hole merger can give
rise to a modest recoil velocity of about $80$ m/s,
assuming that the IMBH is not spinning rapidly. The recoil velocity
magnitude in this case scales as $\propto q^2$
\citep{campanelli07,baker08}, thus implying that the coalescence of
the SMBH with a slowly spinning IMBH more massive than 
$1.5 \times 10^5 \Msun$ can be ruled out based on larger of the observational
constraints, as long as damping of the recoil motion of a remnant SMBH
is inefficient on the time scale of several million years. More stringent 
constraints on the mass of the IMBH, based on the motion of the SMBH 
perpendicular to the Galactic plane, can be placed given the (unknown)
orientation of the orbital plane of the binary before the merger 
in addition to the binary mass ratio and the spin vector of the IMBH.

\subsection{Orientation of the SMBH spin axis}\label{SS_spin}

The nearly perpendicular orientation of the spin axis of the SMBH to
the Galactic disc plane, indicated by the orientation of the observed
gamma-ray bubbles and jets in simulations of \citet{guo11}, implies
that the evolution of the SMBH spin has been determined by accretion
from the Galactic gas disc rather than random accretion events with
isotropic spatial distribution. Such events would include tidal
disruptions of stars and giant molecular clouds triggered by the
satellite inspiral and a merger with the satellite IMBH which orbital
plane in principle may not be aligned with the plane of the Galaxy. It
is thus interesting to consider whether a sequence of such accretion
events can exhibit a cumulative torque on the SMBH sufficient to
re-orient its spin axis, assuming that before the merger with a satellite 
galaxy it was perpendicular to the Galactic plane.

Consider first the effect of episodic gas accretion resulting from
multiple tidal disruption events. \citet{chen09,chen11} show that
three-body interactions between bound stars in a stellar cusp and a
massive binary with properties similar to the IMBH-SMBH considered
here can produce a burst of tidal disruptions, which for a short
period of time ($\sim0.1$~Myr) can exceed the tidal disruption rate
for a single massive black hole by two orders of magnitude, reaching
$\dot{N}\sim 10^{-2}\,{\rm yr^{-1}}$. This implies that in the process
of the IMBH inspiral the SMBH could have disrupted $\sim 10^3$ stars.
A key element in this consideration follows from the finding by
\citet{np98} and \citet{na99} that the orientation of the spin axis of
a SMBH is very sensitive to the angular momentum of the accreted gas:
namely, accretion of a mere few \% in mass of a SMBH can exert torques
that change the direction but not the magnitude of the spin of a black
hole. Because each in a sequence of random accretion events imposes an
infinitesimal change in the orientation of the SMBH spin axis,
collectively they can cause the spin axis to perform a random walk
about its initial orientation. Thus, the magnitude of the effect
scales with the number of disrupted stars and their mass as $\sim
\sqrt{N}\,m_*$. Since this is much less than few percent of $M_1$, the
cumulative effect of tidal disruption events on the orientation of the
spin axis of the SMBH will be negligible. 

This conclusion is reinforced by an additional property of post-tidal
disruption accretion discs: they are compact in size and usually
confined to the region of a size ${\rm few}\times r_t$, where $r_t
\approx r_*\,(M_1/m_*)^{1/3}$ is the tidal disruption radius of a star
and $r_*$ is the stellar radius \citep{rees88}. Such small accretion
discs effectively act as very short lever arms for torques acting on
the spin axis of the SMBH, thus further reducing the efficiency of
this process \citep{np98}. 

Similar conclusions can be reached about the tidally
 disrupted molecular clouds and  gas flows that
plunge towards the SMBH on nearly radial orbits as a consequence of
perturbations excited by the satellite galaxy. In section
\S~\ref{SS_bubble} we estimated that the amount of mass accreted by
the SMBH is $\sim 10^4 \Msun$. A modest mass, combined with the small
circularization radius of the gas accretion disc is insufficient to cause a significant 
change in the SMBH spin orientation. Even ``accretion'' of a spinning 
IMBH is not expected to noticeably
influence the spin orientation of the remnant SMBH. The large mass
ratio of the binary ensures that the final contribution of the IMBH's
spin and orbital angular momentum to the final spin of the SMBH is
small, as long as the pre-merger SMBH has a moderate initial spin, $>{\rm
few}\times 0.1$, in terms of the dimensionless spin parameter
\citep{barausse09}. Hence, coalescence with the IMBH would not
have had a significant effect on the SMBH spin axis orientation.

In summary, the torques from the accretion of tidally disrupted stars,
gas, and the IMBH in the aftermath of the satellite inspiral will be
insufficient to change the orientation of the SMBH spin axis as long
as the SMBH spin is $> {\rm few} \times 0.1$. It follows that the
perpendicular orientation of the spin axis has been set by the
physical processes before the merger with the satellite, and most
likely by the accretion of gas from the Galactic disc.

\section{Conclusions}\label{S_conclusions}

A range of theoretical arguments and observational evidence 
could indicate a satellite infall event within our GC which 
triggered a brief epoch of strong star formation and AGN
activity millions of years ago.  When coupling the newest data -- on
the Fermi bubble and the dearth of late-type stars -- to the
well-established features of the GC such as the cuspy 
early-type stellar
population, a timeline of the recent dynamical events in the galactic
centre emerges.

While the case for a merger of the Milky Way with a satellite galaxy
is not beyond reproach, it is a plausible explanation that naturally accounts
for both the late- and early-type stellar 
distributions and the recent
violent past of Sgr A*.  This event may not be unique in the evolution
of the Milky Way; indeed N-body simulations of the growth of Milky
Way-mass galaxies suggest that the present epoch is rife with mergers
of relic satellite galaxies with the galactic centre, occurring at a
rate of one per few Gyr \citep{diemand07,shb11}. This implies that 
there may have been other bursts of
hypervelocity star ejections, which can seed a population of
``intragroup stars'' farther out in the halo of the
Galaxy. Interestingly, we see tentative evidence in the SDSS archive
for a potential set of very late M giants at $\sim 300\,{\rm kpc}$,
outside the virial radius of our galaxy~\citep{palladinoinprep}. 
Although a followup observation is needed to ensure that these
intragroup candidates are not L dwarfs, if these do prove to be very
distant giants, they may be provide supporting evidence of a previous
minor-merger induced burst of ejected stars $\sim 10^8$ years ago.

Along similar lines, if satellite infall induced activity is common,
then there may be a subset of spiral galaxies which exhibits the signs
of the recent onset of the accretion-powered jets. While the longer
term X- and $\gamma$-ray signatures of jets expanding into the
intragalactic medium may be too faint to observe in galaxies other
than the Milky Way, relatively bright and short lived radio-jets
\citep[$\sim 0.1\,{\rm Myr}$;][]{guo11} may be present in a fraction
of up to $\sim 10^{-4}$ Milky-Way-like spirals, assuming the minor
merger rate cited above. Some of these galaxies may be observed
serendipitously, during the transient phase associated with the onset
of a powerful jet, similar to the case of the previously inactive
galaxy J164449.3+573451 that was recently detected by the Swift
observatory as a powerful source of beamed emission
\citep{burrows11}. If it can be shown that such a sequence of events
occurred in the not so distant past in our Galaxy, it would forever change
the paradigm of the Milky Way as an inactive galaxy with an
underluminous central SMBH.

\section*{Acknowledgments}
We thank our referee for thoughtful suggestions that significantly improved this work. We thank Cole Miller, Melvyn Davies, Jorge Cuadra and Rainer
Sch{\"o}del for insightful discussions. K.H-B., T.B., P.A-S. and
A.S. also acknowledge the hospitality of the Aspen Center for Physics,
where the work was conceived and carried out.  K.H-B. acknowledges the support of a National Science Foundation Career Grant
AST-0847696, and a National Aeronautics and Space Administration Theory grant NNX08AG74G as well as the supercomputing sup-
port of Vanderbilt's Advanced Center for Computation Research and Education, and NASA's Pleiades
and Columbia clusters. Support for
T.B. was provided by the National Aeronautics and Space Administration
through Einstein Postdoctoral Fellowship Award Number PF9-00061 issued
by the Chandra X-ray Observatory Center, which is operated by the
Smithsonian Astrophysical Observatory for and on behalf of the
National Aeronautics Space Administration under contract NAS8-03060.
Support for M.L. was provided by a National Science Foundation Graduate Research Fellowship.



\label{lastpage}


\begin{thebibliography}{99}

\bibitem[\protect\citeauthoryear{Alexander \& Hopman}{2009}]{AlexanderHopman09}
Alexander T., Hopman C., 2009, ApJ, 697, 1861

\bibitem[\protect\citeauthoryear{Amaro-Seoane \& Preto}{2011}]{Amaro-SeoanePreto11}
{Amaro-Seoane} P., {Preto} M., 2011, Classical and Quantum Gravity, 28, 094017

\bibitem[\protect\citeauthoryear{Bahcall \& Wolf}{1976}]{bahcall76}
Bahcall, J.~N. \& Wolf, R.~A.\ 1976, ApJ, 209, 214

\bibitem[\protect\citeauthoryear{Baker et al.}{2008}]{baker08} 
Baker, J.~G., Boggs, W.~D., Centrella, J., Kelly, B.~J., 
McWilliams, S.~T., Miller, M.~C., \& van Meter, J.~R.\ 2008, ApJ, 682, L29

\bibitem[\protect\citeauthoryear{Barausse \& Rezzolla}{2009}]{barausse09} 
Barausse, E., \& Rezzolla, L.\ 2009, ApJ, 704, L40

\bibitem[\protect\citeauthoryear{Barnes \& Hernquist}{1991}]{bh91}
Barnes, J.~E. \& Hernquist, L.\ 1991, ApJ, 370, L65

\bibitem[\protect\citeauthoryear{Barnes \& Hernquist}{1992}]{bh92} Barnes,
J.~E., \& Hernquist, L.\ 1992, ARA\&A, 30, 705 

\bibitem[\protect\citeauthoryear{Barnes \& Hernquist}{1996}]{bh96}
Barnes, J.~E., \& Hernquist, L.\ 1996, ApJ, 471, 115

\bibitem[\protect\citeauthoryear{Bartko et al.}{2010}]{bartko10}
Bartko, H., et al.\ 2010, ApJ, 708, 834

\bibitem[\protect\citeauthoryear{Baumgarte \& Shapiro}{1999}]{baumgarte99} 
Baumgarte, T.~W., \& Shapiro, S.~L.\ 1999, ApJ, 526, 941 

\bibitem[\protect\citeauthoryear{Baumgardt et al.}{2006}]{baum06}
Baumgardt, H., Gualandris, A., \& Portegies Zwart, S.\ 2006, MNRAS,
372, 174

\bibitem[\protect\citeauthoryear{Begelman \& Rees}{1980}]{BegelmanEtAl80}
{Begelman}, M.~C., {Blandford}, R.~D., \& {Rees}, M.~J. 1980, Nature, 287

\bibitem[Begelman et al.(2008)]{begelman08} Begelman, M.~C., 
Rossi, E.~M., \& Armitage, P.~J.\ 2008, MNRAS, 387, 1649 

\bibitem[\protect\citeauthoryear{Bekenstein}{1973}]{bek73}
{Bekenstein}, J.~D. 1973, ApJ, 183, 657

\bibitem[\protect\citeauthoryear{Benson}{2005}]{benson05} 
Benson, A.~J.\ 2005, MNRAS, 358, 551 

\bibitem[\protect\citeauthoryear{Berczik et al.}{2006}]{berzcik06}
Berczik, P., Merritt, D., Spurzem, R., \& Bischof H.~P.	
	 2006, ApJ, 624, 21


\bibitem[\protect\citeauthoryear{Bond et al.}{1991}]{bond91}
Bond, J.~R., Cole, S., Efstathiou, G., \& Kaiser, N.\ 1991, ApJ, 379, 440 

\bibitem[\protect\citeauthoryear{Bonnell \& Rice}{2008}]{bonnell08}
Bonnell, I., \& Rice, W.\ 2008, Science, 321, 1060

\bibitem[\protect\citeauthoryear{Bower}{1991}]{bower91} 
Bower, R.~G.\ 1991, MNRAS, 248, 332 

\bibitem[\protect\citeauthoryear{Boylan-Kolchin et al.}{2004}]{bk04}
Boylan-Kolchin, M., Ma, C.-P., \& Quataert, E.\ 2004, ApJ, 613, L37

\bibitem[\protect\citeauthoryear{Boylan-Kolchin et al.}{2008}]{bk08}
Boylan-Kolchin, M., Ma, C.-P., \& Quataert, E.\ 2008, MNRAS, 383, 93

\bibitem[\protect\citeauthoryear{Bromm \& Loeb}{2003}]{bromm03} 
Bromm, V., \& Loeb, A.\ 2003, ApJ, 596, 34 

\bibitem[\protect\citeauthoryear{Bromm et al.}{2002}]{bromm02} 
Bromm, V., Coppi, P.~S., \& Larson, R.~B.\ 2002, ApJ, 564, 23 

\bibitem[\protect\citeauthoryear{Brown et al.}{2005}]{brown05} 
Brown, W.~R., Geller, M.~J., Kenyon, S.~J., \& Kurtz, M.~J. \ 2005, ApJ, 622, L33

\bibitem[\protect\citeauthoryear{Brown}{2008}]{brown08} 
Brown, W.~R.\ 2008, arXiv:0811.0571

\bibitem[\protect\citeauthoryear{Brown et al.}{2009}]{brown09} 
Brown, W.~R., Geller, M.~J., \& Kenyon, S.~J.\ 2009, ApJ, 690, 1639

\bibitem[\protect\citeauthoryear{Buchholz et al.}{2009}]{buchholz09}
Buchholz, R.~M., Sch{\"o}del, R., \& Eckart, A.\ 2009, A\&A, 499, 483


\bibitem[\protect\citeauthoryear{Bullock et al.}{2001b}]{bullock01b}
Bullock, J.~S., et al.\ 2001, MNRAS, 321, 559

\bibitem[\protect\citeauthoryear{Burrows et al.}{2011}]{burrows11}
Burrows, D.~N., et al.\ 2011 (arXiv:1104.4787)

\bibitem[\protect\citeauthoryear{Callegari et al.}{2011}]{callegari11} 
Callegari, S., Kazantzidis, S., Mayer, L., Colpi, M., Bellovary, J.~M., 
Quinn, T., \& Wadsley, J.\ 2011, ApJ, 729, 85 

\bibitem[\protect\citeauthoryear{Campanelli et al.}{2007}]{campanelli07} 
Campanelli, M., Lousto, C., Zlochower, Y., \& Merritt, D.\ 2007, ApJ, 659, L5

\bibitem[\protect\citeauthoryear{Chandrasekhar}{1943}]{chandra43} 
Chandrasekhar, S.\ 1943, ApJ, 97, 255 

\bibitem[\protect\citeauthoryear{Chen et al.}{2009}]{chen09} 
Chen, X., Madau, P., Sesana, A., \& Liu, F.~K.\ 2009, ApJ, 697, L149 

\bibitem[\protect\citeauthoryear{Chen et al.}{2011}]{chen11} 
Chen, X., Sesana, A., Madau, P., \& Liu, F.~K.\ 2011, ApJ, 729, 13 

\bibitem[\protect\citeauthoryear{Cheng et al.}{2011}]{cheng11} 
Cheng, K.~S., Chernyshov, D.~O., Dogiel, V.~A., Ko, C.~-., \& Ip, W.~-.\
2011, ApJ, 731, L17

\bibitem[\protect\citeauthoryear{Choi et al.}{2009}]{choi09} 
Choi, J.-H., Weinberg, M.~D., \& Katz, N.\ 2009, MNRAS, 400, 1247 

\bibitem[\protect\citeauthoryear{Clark et al.}{2011}]{clark11} 
Clark, P.~C., Glover, S.~C.~O., Klessen, R.~S., \& Bromm, V.\ 2011, 
ApJ, 727, 110 

\bibitem[\protect\citeauthoryear{Colgate}{1967}]{colgate67} 
Colgate, S.~A.\ 1967, ApJ, 150, 163 

\bibitem[\protect\citeauthoryear{Colpi et al.}{1999}]{colpi99} 
Colpi, M., Mayer, L., \& Governato, F.\ 1999, ApJ, 525, 720 

\bibitem[\protect\citeauthoryear{Cox et al.}{2008}]{cox08} 
Cox, T.~J., Jonsson, P., Somerville, R.~S., Primack, J.~R., 
\& Dekel, A.\ 2008, MNRAS, 384, 386 

\bibitem[\protect\citeauthoryear{Crocker et al.}{2011}]{crocker11} 
Crocker, R.~M., Jones, D.~I., Aharonian, F., Law, C.~J., Melia, F., 
Oka, T., \& Ott, J.\ 2011, MNRAS, 413, 763

\bibitem[\protect\citeauthoryear{Cuadra et al.}{2009}]{cuadra09}
Cuadra, J., Armitage, P. J., Alexander, R. D., \& Begelman, M. C. 
2009, MNRAS, 393, 1423

\bibitem[\protect\citeauthoryear{Davis \& Laor}{2011}]{dl11}
Davis, S.~W. \& Laor, A.\ 2011, ApJ, 728, 98

\bibitem[Dehnen \& Binney(1998)]{den98} Dehnen, W., \& Binney, J.\ 1998, MNRAS, 294, 429 


\bibitem[\protect\citeauthoryear{Diemand et al.}{2007}]{diemand07} 
Diemand, J., Kuhlen, M., \& Madau, P.\ 2007, ApJ, 667, 859 

\bibitem[\protect\citeauthoryear{Di Matteo et al.}{2007}]{dimatteo07}
Di Matteo, P., Combes, F., Melchior, A.-L., \& Semelin, B.\ 2007, 
A\&A, 468, 61

\bibitem[\protect\citeauthoryear{Do et al.}{2009}]{do09} Do, T., Ghez,
A.~M., Morris, M.~R., Lu, J.~R., Matthews, K., Yelda, S., \& Larkin,
J.\ 2009, ApJ, 703, 1323

\bibitem[\protect\citeauthoryear{Dobler et al.}{2010}]{dobler10}
Dobler, G., Finkbeiner, D.~P., Cholis, I., Slatyer, T., \& Weiner, N.\
2010, ApJ, 717, 825

\bibitem[\protect\citeauthoryear{Dobler \& Finkbeiner}{2008}]{df08} 
Dobler, G., \& Finkbeiner, D.~P.\ 2008, ApJ, 680, 1222

\bibitem[\protect\citeauthoryear{Dotti et al.}{2007}]{dotti07} 
Dotti, M., Colpi, M., Haardt, F., \& Mayer, L. 2007, MNRAS, 379, 956



\bibitem[\protect\citeauthoryear{Escala et al.}{2005}]{escala05}
Escala, A., Larson, R. B., Coppi, P. S., \& Mardones, D.  2005, ApJ, 630, 152

\bibitem[\protect\citeauthoryear{Faber et al.}{1997}]{faber97} 
Faber, S.~M., et al.\ 1997, AJ, 114, 1771

\bibitem[\protect\citeauthoryear{Ferrarese et al.}{2006}]{ferrarese06}
Ferrarese et al. \ 2006, ApJS, 164, 334

\bibitem[\protect\citeauthoryear{Figer et al.}{2000}]{figer00} 
Figer, D.~F., et al.\ 2000, ApJ, 533, L49

\bibitem[\protect\citeauthoryear{Figer}{2008}]{figer08} 
Figer, D.~F.\ 2008 (arXiv:0803.1619)

\bibitem[\protect\citeauthoryear{Finkbeiner}{2004}]{finke04}
Finkbeiner, D.~P.\ 2004, ApJ, 614, 186

\bibitem[\protect\citeauthoryear{Gao et al.}{2008}]{gao2008}
Gao, L. et al. \ 2008, MNRAS, 387, 536

\bibitem[\protect\citeauthoryear{Genzel et al.}{2003}]{genzel03} 
Genzel, R., et al.\ 2003, ApJ, 594, 812 

\bibitem[\protect\citeauthoryear{Genzel et al.}{2010}]{genzel10}
Genzel, R., Eisenhauer, F., \& Gillessen, S.\ 2010, Reviews of Modern
Physics, 82, 3121


\bibitem[\protect\citeauthoryear{Ghigna et al.}{1998}]{ghigna98} 
Ghigna, S., et al.\ 1998, MNRAS, 300, 146



\bibitem[Gilmore et al.(2002)]{gilmore02} 
Gilmore, G., Wyse, R.~F.~G., \& Norris, J.~E.\ 2002, ApJ, 574, L39 

\bibitem[\protect\citeauthoryear{Gnedin}{2000}]{gnedin00}
Gnedin, N.~Y.\ 2000, ApJ, 542, 535

\bibitem[\protect\citeauthoryear{Gnedin \& Kravtsov}{2006}]{gk06}
Gendin, N.~Y. \& Kravtsov, A.~V.\ 2006, ApJ, 645, 1054

\bibitem[\protect\citeauthoryear{Gnedin \& Ostriker}{1997}]{gnedin97}
Gnedin, O.~Y. \& Ostriker, J.~P. \ 1997, ApJ, 474, 223

\bibitem[\protect\citeauthoryear{Goldreich \& Tremaine}{1979}]{gold79} 
Goldreich, P., \& Tremaine, S.\ 1979, ApJ, 233, 857 

\bibitem[\protect\citeauthoryear{Graham}{2004}]{graham04}
Graham, A.~W.\ 2004, ApJ, 613, L33

\bibitem[\protect\citeauthoryear{Gualandris \& Merritt}{2008}]{gm08}
Gualandris, A., \& Merritt, D.\ 2008, ApJ, 678, 780

\bibitem[\protect\citeauthoryear{Gualandris \& Merritt}{2009}]{gm09}
Gualandris, A., \& Merritt, D.\ 2009, ApJ, 705, 361

\bibitem[\protect\citeauthoryear{Gualandris, Gillessen \& Merritt}{2010}]{ggm10}
Gualandris, A., Gillessen, S., \& Merritt, D.\ 2010, MNRAS, 409, 1146

\bibitem[\protect\citeauthoryear{Gualandris \& Merritt}{2012}]{gm12}
Gualandris, A., \& Merritt, D.\ 2012, ApJ, 744, 74

\bibitem[\protect\citeauthoryear{Guo \& Mathews}{2011}]{guo11}
Guo, F. \& Mathews, W. G.\ 2011, Draft

\bibitem[\protect\citeauthoryear{Hansen \& Milosavljevi{\'c}}{2003}]{hm03} 
Hansen, B.~M.~S., \& Milosavljevi{\'c}, M.\ 2003, ApJ, 593, L77

\bibitem[\protect\citeauthoryear{Hernquist}{1990}]{hernquist90}
Hernquist, L.\ 1990, ApJ, 356, 359


\bibitem[\protect\citeauthoryear{Holley-Bockelmann et al.}{2010}]{KHB10} 
Holley-Bockelmann, K., Micic, M., Sigurdsson, S., 
\& Rubbo, L.~J.\ 2010, ApJ, 713, 1016 

\bibitem[\protect\citeauthoryear{Hopkins et al.}{2008}]{hopkins08}
Hopkins, P.~F. et al.\ 2008, ApJ, 688, 757

\bibitem[\protect\citeauthoryear{Hopkins et al.}{2009}]{hopkins09}
Hopkins, P.~F., Cox, T.~J., Younger, J.~D., \& Hernquist, L.\ 2009, ApJ, 691, 1168


\bibitem[\protect\citeauthoryear{Hopkins \& Hernquist}{2010}]{hh10}
Hopkins \& Hernquist \ 2010, MNRAS, 407, 447

\bibitem[\protect\citeauthoryear{Hopkins \& Quataert}{2010}]{hq10}
Hopkins, P.~F \& Quataert, E.\ 2010, MNRAS, 407, 1529

\bibitem[\protect\citeauthoryear{Hopkins \& Quataert}{2011}]{hq11}
Hopkins, P.~F \& Quataert, E.\ 2011, MNRAS, 415, 1027

\bibitem[\protect\citeauthoryear{Inui et al.}{2009}]{inui09} 
Inui, T., Koyama, K., Matsumoto, H., \& Tsuru, T.~G.\ 2009, PASJ, 61, S241

\bibitem[\protect\citeauthoryear{Irrgang et al.}{2010}]{irrgang10}
Irrgang, A., Przybilla, N., Heber, U., Nieva, M.~F., \& Schuh, S.\
2010, ApJ, 711, 138

\bibitem[\protect\citeauthoryear{Jogee et al.}{2005}]{jogee05}
Jogee, S., Scoville, N., \& Kenny, J.~D.~P. \ 2005, ApJ, 630, 837

\bibitem[\protect\citeauthoryear{Just \& Pe{\~n}arrubia}{2005}]{just05} 
Just, A., \& Pe{\~n}arrubia, J.\ 2005, A\&A, 431, 861 

\bibitem[\protect\citeauthoryear{Khan et al.}{2011}]{khan11} 
Khan, F.~M., Just, A., \& Merritt, D.\ 2011, ApJ, 732, 89

\bibitem[Khochfar \& Burkert(2006)]{khochfar06} 
Khochfar, S., \& Burkert, A.\ 2006, A\&A, 445, 403 


\bibitem[\protect\citeauthoryear{Klypin et al.}{2011}]{klypin11}
Klypin, A.~A., Trujillo-Gomez, S., \& Primack, J. \ 2011, ApJ, 740, 102

\bibitem[\protect\citeauthoryear{Kormendy \& Bender}{2009}]{kb09}
Kormendy, J., \& Bender, R.\ 2009, ApJ, 691, L142

\bibitem[\protect\citeauthoryear{Krabbe et al.}{1995}]{krabbe95}
Krabbe, A., et al.\ 1995, ApJ, 447, L95


\bibitem[\protect\citeauthoryear{Lacey \& Cole}{1993}]{lc93} 
Lacey, C., \& Cole, S.\ 1993, MNRAS, 262, 627 

\bibitem[\protect\citeauthoryear{Lindqvist et al.}{1992}]{lindqvist92} 
Lindqvist, M., Habing, H.~J., \& Winnberg, A.\ 1992, A\&A, 259, 118 

\bibitem[Loeb \& Rasio(1994)]{loeb94} 
Loeb, A., \& Rasio, F.~A.\ 1994, ApJ, 432, 52 

\bibitem[\protect\citeauthoryear{Loose et al.}{1982}]{loose82}
Loose, H.~H., Kruegel, E., \& Tutukov, A. \ 1982, A\&A, 105, 342

\bibitem[Madau \& Rees(2001)]{madau01} 
Madau, P., \& Rees, M.~J.\ 2001, ApJ, 551, L27 

\bibitem[\protect\citeauthoryear{Mapelli et al.}{2012}]{mapelli12}
Mapelli, M., Hayfield, T., Mayer, L., \& Wadsley, J.\ 2012, ApJ, 749, 168


\bibitem[\protect\citeauthoryear{Matsubayashi et al.}{2007}]{matsu07}
Matsubayashi, T., Makino, J., \& Ebisuzaki, T.\ 2007, ApJ, 656, 879

\bibitem[\protect\citeauthoryear{McBride et al.}{2009}]{mcbride09}
McBride, J., Fakhouri, O., \& Ma, C.\ 2009, MNRAS, 398, 1858

\bibitem[\protect\citeauthoryear{Merritt}{2006}]{merritt06} 
Merritt, D.\ 2006, ApJ, 648, 976

\bibitem[\protect\citeauthoryear{Merritt}{2010}]{merritt10} 
Merritt, D.\ 2010, ApJ, 718, 739

\bibitem[\protect\citeauthoryear{Merritt \& Cruz}{2001}]{merritt01}
Merritt, D. \& Cruz, F. \ 2001, ApJ, 551, L41

\bibitem[\protect\citeauthoryear{Merritt \& Poon}{2004}]{merritt04} 
Merritt, D.\& Poon, M. Y. \ 2004, ApJ, 606, 788

\bibitem[\protect\citeauthoryear{Merritt, Gualandris \& Mikkola}{2009}]{merritt09} 
Merritt, D., Gualandris, A., \& Mikkola, S. \ 2009, ApJ, 693, L35

\bibitem[\protect\citeauthoryear{Micic et al.}{2011}]{micic11} 
Micic, M., Holley-Bockelmann, K., \& Sigurdsson, S.\ 2011, MNRAS, 414, 1127 

\bibitem[\protect\citeauthoryear{Mihos \& Hernquist}{1996}]{mh96}
Mihos, J.~C. \& Hernquist, L.\ 1996, ApJ, 464, 641

\bibitem[\protect\citeauthoryear{Miller}{2002}]{Miller02}
{Miller}, M.~C. 2002, ApJ, 581, 438

\bibitem[\protect\citeauthoryear{Milosavljevi{\'c} \& Merritt}{2001}]{mm01} 
Milosavljevi{\'c}, M., \& Merritt, D.\ 2001, ApJ, 563, 34

\bibitem[\protect\citeauthoryear{Miyamoto \& Nagai}{1975}]{mn75}
Miyamoto, M. \& Nagai, R.\ 1975 , PASJ, 27, 533

\bibitem[\protect\citeauthoryear{Molinari et al.}{2011}]{molinari11} 
Molinari, S., Bally, J., Noriega-Crespo, A., et al.\ 2011, ApJ, 735, L33 

\bibitem[\protect\citeauthoryear{Noguchi}{1988}]{noguchi88}
Noguchi, M.\ 1988, A\&A, 203, 259

\bibitem[\protect\citeauthoryear{Natarajan \& Armitage}{1999}]{na99}
Natarajan, P., \& Armitage, P.~J.\ 1999, MNRAS, 309, 961

\bibitem[\protect\citeauthoryear{Natarajan \& Pringle}{1998}]{np98}
Natarajan, P., \& Pringle, J.~E.\ 1998, ApJ, 506, L97

\bibitem[\protect\citeauthoryear{Navaro, Frenk, \& White}{1997}]{nfw97}
Navaro, J.~F., Frenk, C.~S., \& White, S.~D.~M.\ 1997, ApJ, 490, 493

\bibitem[\protect\citeauthoryear{Ostriker}{1999}]{ostriker99} 
Ostriker, E.~C.\ 1999, ApJ, 513, 252 



\bibitem[Palladino et al.(2012)]{palladinoinprep} Palladino, L.~E., 
Holley-Bockelmann, K., Morrison, H., et al.\ 2012, AJ, 143, 128 





\bibitem[\protect\citeauthoryear{Papaloizou \& Pringle}{1977}]{papa77} 
Papaloizou, J., \& Pringle, J.~E.\ 1977, MNRAS, 181, 441 


\bibitem[\protect\citeauthoryear{Parkinson et al.}{2008}]{pch08}
Parkinson, H., Cole, S., \& Helly, J.\ 2008, MNRAS, 383, 557

\bibitem[\protect\citeauthoryear{Paumard et al.}{2006}]{paumard06}
Paumard, T., et al.\ 2006, ApJ, 643, 1011

\bibitem[\protect\citeauthoryear{Peres}{1962}]{per62} 
Peres, A. 1962, Phys. Rev., 128, 2471

\bibitem[\protect\citeauthoryear{Perets \& Alexander}{2008}]{perets08} 
Perets, H. B., \& Alexander, T. 2008, ApJ, 677, 146

\bibitem[\protect\citeauthoryear{Perets \& Gualandris}{2010}]{perets10} 
Perets, H. B., \& Gualandris, A.\ 2010, ApJ, 719, 220

\bibitem[\protect\citeauthoryear{Pe{\~n}arrubia et al.}{2004}]{pen04} 
Pe{\~n}arrubia, J., Just, A., \& Kroupa, P.\ 2004, MNRAS, 349, 747 

\bibitem[\protect\citeauthoryear{Peters \& Mathews}{1963}]{pm63} 
Peters, P. C., \& Mathews, J. 1963, Phys. Rev., 131, 435

\bibitem[\protect\citeauthoryear{Ponti et al.}{2010}]{ponti10} 
Ponti, G., Terrier, R., Goldwurm, A., Belanger, G., \& Trap, G.\ 2010, ApJ,
714, 732

\bibitem[\protect\citeauthoryear{Portegies Zwart et al.}{2006}]{pz06}
Portegies Zwart, S.~F., Baumgardt, H., McMillan, S.~L.~W., Makino, J.,
Hut, P., \& Ebisuzaki, T.\ 2006, ApJ, 641, 319

\bibitem[\protect\citeauthoryear{Prada et al.}{2011}]{prada11}
Prada, F., Klypin, A.~A., Cuesta, A.~J., Betancort-Rijo, J.~E., 
\& Primack, J.\ 2011, eprint arXiv:1104.5130

\bibitem[\protect\citeauthoryear{Preto \& Amaro-Seoane}{2010}]{PretoAmaroSeoane10}
{Preto} M., {Amaro-Seoane} P., 2010, ApJ, 708, L42

\bibitem[\protect\citeauthoryear{Preto et al.}{2011}]{preto11} 
Preto, M., Berentzen, I., Berczik, P., \& Spurzem, R.\ 2011, ApJ, 732, L26

\bibitem[\protect\citeauthoryear{Quinlan \& Shapiro}{1987}]{quinlan87} 
Quinlan, G.~D., \& Shapiro, S.~L.\ 1987, ApJ, 321, 199 

\bibitem[\protect\citeauthoryear{Quinlan}{1996}]{quinlan96}
Quinlan, G.~D.\ 1996, New Astronomy, 1, 35

\bibitem[\protect\citeauthoryear{Quinn \& Goodman}{1986}]{qg86}
Quinn, P. J. \& Goodman, J.\ 1986, ApJ, 309, 472Q

\bibitem[\protect\citeauthoryear{Quinn, Hernquist, \& Fullagar}{1993}]{qhf93}
Quinn, P.~J., Hernquist, L., \& Fullagar, D.~P.\ 1993, ApJ, 403, 74

\bibitem[\protect\citeauthoryear{Rees}{1988}]{rees88} 
Rees, M.~J.\ 1988, Nature, 333, 523

\bibitem[\protect\citeauthoryear{Reid \& Brunthaler}{2004}]{rb04}
Reid, M.~J., \& Brunthaler, A.\ 2004, ApJ, 616, 872

\bibitem[\protect\citeauthoryear{Ricotti \& Gnedin}{2005}]{rg05}
Ricotti, M. \& Gnedin, N.~Y.\ 2005, ApJ, 629, 259

\bibitem[\protect\citeauthoryear{Ricotti et al.}{2008}]{ricotti08} 
Ricotti, M., Gnedin, N.~Y., \& Shull, J.~M.\ 2008, ApJ, 685, 21 

\bibitem[\protect\citeauthoryear{Rownd \& Young}{1999}]{ry99}
Rownd, B.~K. \& Young, J.~S.\ 1999, ApJ, 118, 670

\bibitem[\protect\citeauthoryear{S{\'a}nchez-Salcedo \& Brandenburg}{1999}]{sanchez99} 
S{\'a}nchez-Salcedo, F.~J., \& Brandenburg, A.\ 1999, ApJ, 522, L35 

\bibitem[\protect\citeauthoryear{Sch\"{o}del et al.}{2009}]{schodel09}
Sch\"{o}del, R., Merrit, D., \& Eckart, A.\ 2009, A\&A, 502, 91

\bibitem[\protect\citeauthoryear{Sellwood, Nelson, \& Tremaine}{1998}]{snt98}
Sellwood, J.~A., Nelson, R.~W., \& Tremaine, S.\ 1998, ApJ, 506, 590

\bibitem[\protect\citeauthoryear{Sesana et al.}{2007}]{sesanaetal07}
Sesana, A., Haardt, F., \& Madau, P.\ 2007, MNRAS, 379, 45

\bibitem[\protect\citeauthoryear{Sesana et al.}{2008}]{sesana08}
Sesana, A., Haardt, F., \& Madau, P.\ 2008, ApJ, 686, 432

\bibitem[\protect\citeauthoryear{Sesana}{2010}]{sesana10} 
Sesana, A.\ 2010, ApJ, 719, 851

\bibitem[\protect\citeauthoryear{Shakura \& Sunyaev}{1973}]{ss73}
Shakura, N.~I. \& Sunyaev, R.~A.\ 1973, A\&A, 24, 337

\bibitem[\protect\citeauthoryear{Simon \& Geha}{2007}]{sg07}
Simon, J.~D. \& Geha, M.\ 2007, ApJ, 670, 313

\bibitem[Sinha \& Holley-Bockelmann(2012)]{shb11} Sinha, M., \& Holley-Bockelmann, K.\ 2012, ApJ, 751, 17 


\bibitem[\protect\citeauthoryear{Snowden et al.}{1997}]{snowden97}
Snowden, S.~L., et al.\ 1997, ApJ, 485, 125

\bibitem[\protect\citeauthoryear{Stolte et al.}{2008}]{StolteEtAl08}
Stolte, A., Ghez, A.~M., Morris, M., Lu, J.~R., Brandner, W., \& Matthews, K. 2008, ApJ, 675, 1278
  
\bibitem[\protect\citeauthoryear{Strigari et al.}{2008}]{strigari08}
Strigari, L.~E., Bullock, J.~S., Kaplinghat, M., Simon, J.~D., Geha, M., 
William, B., \& Walker, M.~G.\ 2008, ApJ, 454, 1096

\bibitem[\protect\citeauthoryear{Su et al.}{2010}]{su10} 
Su, M., Slatyer, T.~R., \& Finkbeiner, D.~P.\ 2010, ApJ, 724, 1044

\bibitem[\protect\citeauthoryear{{\v S}ubr et al.}{2009}]{subr09} 
{\v S}ubr, L.~., Schovancov{\'a}, J., \& Kroupa, P.\ 2009, A\&A, 496, 695

\bibitem[\protect\citeauthoryear{Stark et al.}{2004}]{stark04} 
Stark, A.~A., Martin, C.~L., Walsh, W.~M., Xiao, K., Lane, A.~P., 
\& Walker, C.~K.\ 2004, ApJ, 614, L41 

\bibitem[\protect\citeauthoryear{Taylor \& Babul}{2001}]{tb01}
Taylor, J.~E. \& Babul, A.\ 2001, ApJ, 559, 716

\bibitem[\protect\citeauthoryear{Terrier et al.}{2010}]{terrier10}
Terrier, R., et al.\ 2010, ApJ, 719, 143

\bibitem[\protect\citeauthoryear{Tillich et al.}{2009}]{tillich09}
Tillich, A., Przybilla, N., Scholz, R.-D., \& Heber, U.\ 2009, A\&A,
507, L37

\bibitem[\protect\citeauthoryear{Tormen et al.}{1997}]{tormen97} 
Tormen, G., Bouchet, F.~R., \& White, S.~D.~M.\ 1997, MNRAS, 286, 865 

\bibitem[\protect\citeauthoryear{Toth \& Ostriker}{1992}]{toth92} 
Toth, G., \& Ostriker, J.~P.\ 1992, ApJ, 389, 5 


\bibitem[\protect\citeauthoryear{Vel\'{a}zquez \& White}{1999}]{vw99}
Vel\'{a}zquez, H. \& White, S.~D.~M.\ 1999, MNRAS, 304, 254

\bibitem[\protect\citeauthoryear{Vesperini \& Weinberg}{2000}]{vw00}
Vesperini, E. \& Weinberg, M.\ 2000, ApJ, 534, 598

\bibitem[\protect\citeauthoryear{Walker et al.}{1996}]{walker96} 
Walker, I.~R., Mihos, J.~C., \& Hernquist, L.\ 1996, ApJ, 460, 121 

\bibitem[\protect\citeauthoryear{Wang et al.}{2005}]{wang05} 
Wang, H.~Y., Jing, Y.~P., Mao, S., \& Kang, X.\ 2005, MNRAS, 364, 424 

\bibitem[\protect\citeauthoryear{Weinberg}{1997}]{weinberg97} 
Weinberg, M.~D.\ 1997, ApJ, 478, 435 

\bibitem[\protect\citeauthoryear{Wetzel}{2011}]{wetzel11} 
Wetzel, A.~R.\ 2011, MNRAS, 412, 49 

\bibitem[\protect\citeauthoryear{Widrow \& Dubinski}{2005}]{wid05} 
Widrow, L.~M., \& Dubinski, J.\ 2005, ApJ, 631, 838 

\bibitem[\protect\citeauthoryear{Wise \& Abel}{2008}]{wa08}
Wise, J.~H. \& Abel, T.\ 2008, ApJ, 684, 1

\bibitem[\protect\citeauthoryear{Yelda et al.}{2010}]{yelda10} 
Yelda, S., Lu, J.~R., Ghez, A.~M., Clarkson, W., Anderson, J., 
Do, T., \& Matthews, K.\ 2010, ApJ, 725, 331 

\bibitem[\protect\citeauthoryear{Zentner et al.}{2005}]{zentner05} 
Zentner, A.~R., Kravtsov, A.~V., Gnedin, O.~Y., \& 
Klypin, A.~A.\ 2005, ApJ, 629, 219 

\bibitem[\protect\citeauthoryear{Zhao et al.}{2003}]{zhao03}
Zhao, D.~H., Jing, Y.~P., Mo, H.~J., \& B\"{o}rner, G.\ 2003, ApJ, 597, L9

\bibitem[\protect\citeauthoryear{Zubovas et al.}{2011}]{zubovas11}
Zubovas, K., King, A.~R., \& Nayakshin, S.\ 2011, MNRAS, L274

\end{thebibliography}
\end{document}